# Understanding Microplasmas

**J. Winter**, J. Benedikt, M. Böke, D. Ellerweg, T. Hemke, N. Knake, T. Mussenbrock, B. Niermann, D. Schröder, V. Schulz-von der Gathen, A. von Keudell

*Institute for Experimental Physics II, Research Department "Plasmas with Complex Interactions", Ruhr- Universität Bochum, Germany*

## 0. Abstract

Microplasmas are operated around atmospheric pressure exhibiting pronounced non-equilibrium characteristics, i.e. they possess energetic electrons while ions and neutrals remain cold. They have gained significant interest due to their large application potential e.g. in the biomedical, surface modification and light source areas, just to name a few. Many different configurations are in use. Their understanding and quantification is mandatory for further progress in applications. As an example, we discuss the case of a micro plasma jet operated in $He/O_2$ in an ambient air environment as prototype micro plasma for various applications. We present experimental and modeling work encompassing the entire micro plasma system starting from gas injection via the bulk plasma up to the contact with a target.

## 1. Introduction

Microplasmas operated at atmospheric pressure have recently gained very high attention. A series of review papers and topical issues of leading scientific journals have discussed their technological promises as well as their scientific challenges [1, 2, 3, 4].

Microplasmas are characterized by high electron densities in small volumes. The characteristic length scale for the charged particle dynamics, the Debye length, contracts below a value of 1 μm. The mean free path of particles and photons in these plasmas becomes rather small. Energetic electrons, however, may have mean free paths of the order of the discharge vessel, leading to non local plasma heating. The surface-to-volume ratio is extreme, yielding very high power densities.

Research on microscopic plasmas nowadays is aiming at the development and optimization of plasmas for light emission or for the production of reactive species from a molecular precursor gas. Light emission is exploited in display panels or in short arcs as used for extreme UV lithography. Reactive micro plasmas have been developed for surface treatment of materials ranging from polymers up to living tissue in the emerging field of plasma medicine. Also depositing gases are used for the deposition of functional organic or inorganic coatings and thin films. An analysis of the present state of the art clearly shows that the basic understanding of micro plasmas is often limping behind, impeding their scaling and is impeding, in particular in the bio-medical area, the approval of processes by authorities.

There are many different micro plasma configurations in use ranging from micro jets [5], micro hollow cathode arrangements [6] to arrays of integrated structures [7, 8] with excitations ranging from the microwave range [9], radio frequency [10, 11], pulsed high voltage [12], to continuous DC power [13].

Micro jets are particularly well suited for surface treatment with reactive species, since the forced gas flow through the jet guides these species to the surface. The non-equilibrium character renders them particularly useful for the localized treatment of thermo labile materials, in particular tissue and living cells. This makes them prominent candidates for applications in the medical environment – plasma medicine is a rapidly developing field [14]. The remote character of the treatment implies that due to quenching, recombination and plasma chemical reactions with the ambient surrounding of the jet, the species distribution in the discharge region as such and in the effluent may differ vastly. Species may disappear rapidly and new ones may be created. It is evident, that simple "before plasma treatment – after plasma treatment" considerations are hardly useful.

It is thus important to understand not only the plasma generation, its energy flow and its dynamics and stability but also to consider carefully all steps involved until the species really reach the substrate to be treated. This is a most challenging task for diagnostics and modeling. Absolute measurements of discharge parameters and fluxes are needed to perform ultimate benchmarks and tests for the falsification or verification of current hypotheses and theories. Without quantitative data further progress is very much hampered and interpretation of the complex phenomena involving micro plasma applications will remain in the realm of tell stories.

A particularly popular class of micro plasmas is radio-frequency driven plasma jets, originally proposed by Selwyn and co-workers in 1998 [15]. Many groups have investigated their prospect for surface modification; the studied processes include etching of tungsten, deposition and etching of silicon oxide, and cleaning of thermo labile surfaces from contaminants [16; 17; 18; 19; 20; 21], for example. They are also popular systems for the investigation of plasmas interacting with living tissue.

As a prototype for a micro plasma jet this paper focuses on measurements and modeling done with the coplanar micro scale atmospheric pressure plasma jet (μ-APPJ). We will present quantitative data on the kinetics of oxygen species in a RF excited jet operated in He/$O_2$ and its interaction with surfaces, discuss the problem of energy flow in the jet by metastable He and Ar atoms which is associated with the stability of the micro jet and the problem of impurity intrusion into jets when operated in ambient atmosphere.

## 2. The μ-APPJ

The micro plasma jet ($\mu$-APPJ) used for the experiments is a capacitively coupled radio frequency discharge (13.56 MHz, 15W RF power) designed for optimized optical diagnostic access [11] (Fig.1). The coplanar geometry and an electrode width of about 1mm provide a discharge profile of 1mm$^2$ for localized surface treatment at low gas consumption. It is operated in a homogeneous glow mode with a noble gas flow (typically 1.4 slm He) containing a small admixture of molecular oxygen (typically 0.5%). For some investigations the jet was prolonged in its longitudinal direction (x-axis) by a non powered section made from isolating materials to avoid immediate contact with ambient air and guide to the gas flow.

One of the electrodes is grounded, the other one is powered with the radio frequency through an impedance matching network. The electric field between the electrodes causes a breakdown in the gas and produces a plasma with electron temperature and density of about 1 to 2 eV and $10^{10}$cm$^{-3}$, respectively [22, 23]. Atoms and molecules in the feed gas become excited, dissociated or ionized by electron impact. Since the electrons are not in thermal equilibrium with the ions and neutrals, the gas temperature remains a few tens of K above room temperature [24]. In order to vary the (background) pressure of the jet and to control the gas composition around the jet, the whole system may be placed inside a vacuum chamber with an additional gas controlling system. Hence, studies of the interdependency between the background gas and the jet with its characteristics can be carried out.

## 3. Absolute measurements and modeling of reactive O species

Ground state atomic oxygen densities were measured by two-photon absorption laser-induced fluorescence spectroscopy (TALIF) [25] providing space resolved density maps. In TALIF spectroscopy ground state atomic oxygen is excited into a higher state by simultaneous absorption of two ultraviolet photons (to avoid the use of single photons of the vacuum ultraviolet wavelength range for overcoming the large energy gap). The fluorescence radiation emitted from that higher state is proportional to the ground state atomic oxygen density. At atmospheric pressure several particular effects have to be taken into account. The most important ones are various saturation effects depending on laser power, such as laser induced additional particle generation. To avoid this, the laser intensity has to be kept as low as possible. Due to the nature of two-photon excitation, the fluorescence light depends quadratically on the exciting laser intensity. In the case of saturation effects as mentioned above, this quadratic dependence is disturbed. To exclude these saturation effects it has to be ensured that all TALIF measurements are performed in the regime of quadratic dependence. Another feature of particular importance is the non-radiative de-excitation of the excited state by collisional quenching which reduces the natural lifetime $\tau$. These quenching rates have to be known for the dominant colliding species in order to perform a quantitative analysis of the fluorescence data, e.g. to obtain density maps. They were either taken from the literature [26] or measured for comparable systems up to atmospheric pressure [25]. It has to be considered that for atomic oxygen, both the ground state and the two-photon excited state are triplet states. The resulting line broadening has to be taken into account for evaluation together with Doppler and pressure line broadening. To yield absolute concentrations the TALIF measurements must be calibrated. We applied a calibration method based on comparative TALIF measurements using xenon as a noble gas reference with a two-photon resonance, spectrally close to that of the atomic oxygen to be quantified [26]. A further advantage of this method is the possibility to perform localized calibrations for the entire observation field. For details see ref [25].

2-dimensional distributions of atomic oxygen have been measured from the gas inlet, through the electrode region of the jet device to the end of the electrodes, and finally into the effluent up to a surface positioned in the effluent (see section 5).

Fig. 2 shows the absolute oxygen densities along the x- axis of the jet, with the measuring spot always in the middle between electrodes. In agreement with modelling, the maximum O density in the jet is achieved for an admixture of 0.6% $O_2$ to He.

As can also be seen in Fig. 2, the atomic densities in the core region reach a plateau value after distances of several millimeters or ascent times of about 60 µs [25; 27]. This increase can be understood under the simplifying assumptions of a constant production term (electron dissociation) and a loss term proportional to the oxygen density.
An initially surprising result was an increase of the atomic density to about $4\times10^{15}$cm$^{-3}$ with increasing gas velocity or mass flow (Fig. 2). One possible explanation, the influence of impurity intrusion from the outside, is discussed in section 4. The puzzle is not fully resolved yet. In agreement with modelling [28], the maximum O density in the jet is achieved for an admixture of 0.6% $O_2$ to He.

Complementary measurements of reactive oxygen species were performed by calibrated molecular beam mass spectroscopy in the effluent (see Fig. 3). The recently developed MBMS diagnostics [29], comprises a molecular beam mass spectrometer system with three pumping stages and with a beam chopper with rotating skimmer. This system provides high signal intensity and an excellent beam-to-background ratio. The central idea is to modulate the extracted beam using a chopper in the first pumping stage. Thereby, the pressure in the second stage can be very low, and the time modulated beam can be used to separate beam from background. This MBMS system was used to characterize the µ-APPJ operated in He/$O_2$ and gas mixtures. Atomic oxygen and ozone ($O_3$) in the effluent of the µ-APPJ have been quantified as a function of applied power, $O_2$ admixture, and distance to the microjet [30].

O densities in the effluent agree very well with the TALIF data regarding the trend. Absolute numbers differ by about a factor of 0.27 which is likely to be due to the presence of the surface of the mass spectrometer. It is also seen that significant amounts of $O_3$ are produced by reactions of O with $O_2$ whereby one source of O remote from the nozzle may be due to VUV radiation propagating along the He beam leaving the nozzle.

By applying a 2-dimensional model, generation and transport of relevant species from the discharge region to the effluent of the plasma jet were studied in an ambient helium atmosphere. In detail, the fluid code nonPDPSIM developed by M.Kushner and co-workers was used to simulate the plasma jet. We have taken into account the physics and chemistry of charged particles: electrons, positive and negative ions, and also excited as well as ground state neutrals. For all species the continuity equations were simultaneously solved. The particle fluxes have been calculated from the momentum balances in the drift-diffusion approximation evaluated in the local center-of-mass system. For the electron fluid, the energy balance equation has been additionally solved. It takes into account Ohmic heating and the energy losses due to elastic and inelastic interaction with neutrals and ions as well as heat conduction.

To capture the non-Maxwellian behavior of the electrons, all electronic transport coefficients (mobility, diffusion constant, and thermal conductivity) as well as the electronic rate coefficients have been calculated by the local mean energy method: A zero-dimensional Boltzmann equation for the electron energy distribution and the transport and rate coefficients has been solved for the locally applicable gas composition and various values of the electrical field. The tabulated data have then been consulted in dependence of the dynamically calculated electron temperature. The plasma equations have been coupled with a modified version of the compressible Navier-Stokes equations, which have been solved for the gas density, the mean velocity, and the gas temperature. The contributions to the energy equation from Joule heating include only ion contributions. The heat transfer from the electrons is included as a collisional change of enthalpy. Finally, the electric potential has been calculated from Poisson's equation. Whereas the charge density stems from the charged particles in the plasma domain and from the charges at the surfaces, the surface charges are governed by a separate balance equation. The dynamical equations have been complemented by an appropriate set of boundary conditions. Electrically, the walls are either powered or grounded. With respect to the particle flow, they are either solid, or represent inlets or outlets. The inlet flow is specified to a given flux, while the outlet flow is adjusted to maintain the pressure. Finally, it is worth mentioning that the actual implementation of the equations poses some difficulties due to the vast differences in the time scales of the dynamics of the plasma and the neutrals. These difficulties have been overcome by the methods of time-slicing and subcycling.

We have found a pronounced difference in the behavior of charged particles and neutrals. The charged particles and predominantly $O_2^+$, $O^+$, $O^-$, $O_2^-$ (see Fig. 5) are governed by a "fast" chemistry: In the jet, their production is local in x and their losses are dominated by drift and diffusion to the electrodes. In the effluent, they are virtually absent. The neutrals, in contrast, obey a "slow" chemistry. The reactive oxygen species ground-state O, $O_3$ and $O_2$ ($^1\Delta$) build up along the x-axis of the jet and are then transported by the gas flow far into the effluent (see Fig. 4). As shown in Fig. 3 this is confirmed excellently by the data absolutely measured by TALIF and MBMS [30; 31]. The simulation results indicate that the influence of the jet on the environment is primarily of chemical nature. Even in the jet, the density of the activated oxygen species is four orders of magnitude higher than the maximum density of charged particles, and in the effluent charged particles vanish almost completely. In particular ozone with a very long lifetime, ground-state atomic oxygen and $O_2$ ($^1\Delta$) are able to react with surfaces over a long distance. Excited atomic oxygen species, in contrast, play a relatively minor role due to their lower concentration.

## 4. He metastables - the role of impurity intrusion

The absolute concentrations of He* and Ar* metastables were measured using tunable laser diode absorption spectroscopy (TDLAS). The μ-APPJ was operated in a pulsed mode in order to apply lock-in techniques and to study lifetimes of species in the afterglow (pulse frequency: 4 kHz, duty cycle: 50%). 2-dimensional maps of the metastables' concentrations and their lifetimes were resolved for various plasma conditions. A detailed description of the experimental set up can be found in [32]. Only a rough description is given here.

For the TDLAS measurements, two commercial diode laser (DL) systems were used. This was an external cavity for the 811.5 nm argon line and a DL for the 1083 nm helium line, both with a line width small compared to that of the absorption lines. Figure 8 shows a sketch of the experimental setup. The laser beam from the DL passed through two beam splitters. A part of the beam was guided to a Fabry- Perot interferometer (1 GHz free spectral range) and another part through a low pressure reference cell to perform the calibration of the laser frequency. An attenuated part of the beam transmitted through the first beam splitter was guided into the jet perpendicular to the quartz windows that confined the gas channel. With a spot size of less than 150 *μ*m at the focal point this provides a good spatial resolution in the *x-z*-plane of the plasma volume. The absorption length is limited to 1 mm, the distance between the windows.

Fig. 8 shows a 2D-map of the He$^*$ ($^3S_1$) density in the discharge volume. Both, horizontal and vertical axis show the area between the electrodes. The map covers 2000 reading points (40 vertical x 50 horizontal) of the absorption signal. The measurement shown here was made in a pure He discharge. The density of metastables depends sensitively on the intrusion of impurities (see below). Concerning the effluent region behind the electrodes, no metastables can be detected even at high power. This is comprehensible regarding the short lifetimes of He*. With a feed gas flow of 2 slm in the discharge channel and a metastable lifetime usually much less than 6 μs, the metastable density is expected to vanish after about 300 μm (3 times their decay length) without further excitation. Local changes in the density profile along the horizontal axis are likely to be due to local irregularities in the electrode surface.

A density profile of the metastable distribution between the electrodes is shown in Fig. 9 for both argon and helium metastables. The measurement of the Ar* shown here was made in a He discharge with 3% Ar admixture, the He* density was measured in pure helium. The graphs show a steep rise in front of the electrode surface and a decline in the bulk of the plasma after passing the maximum. This behavior is consistent with our understanding of the sheath structure in an RF excited discharge. Directly in front of the electrodes, the metastable density is low since the electron density is too low for an efficient excitation and ionization of the ground state atoms. Metastables reach their highest densities some 100 μm away from the electrode, where in the negative glow area most excitation and ionization processes occur. In the bulk, the electron temperature is low, yielding a lower metastable production rate. The Ar* density in the bulk is significantly higher than the He* density due to the lower excitation threshold. Electrons in the bulk are still fast enough to cause Ar* production. In the Ar* profile, a double peak structure can be observed in front of both electrodes. The first peak is more narrow and seems to coincide with the maximum of the He* density. The second is broader and about 200 μm deeper into the discharge. This structure suggests two different production mechanisms for the Ar* species (electron excitation and Penning collisions with He*). The metastables are a very important reservoir of potential energy, which is converted into ionization by pooling reactions. This is clearly seen in the excitation dynamics and location of molecular ions, e.g. $N_2^+$ in the case of $N_2$ admixture. The profiles are well reproduced by modelling.

The influence of different gas admixtures on the density of He* and their life time has been investigated in detail (Fig. 10 left) shows that already a few ppm of Ar change the He* density significantly. This effect is even more pronounced for the admixture of molecular gases, e.g. for $N_2$ or $O_2$, due to quenching of these states by Penning collisions. The same effect leads to a reduced lifetime of the metastable states. The measurements shown in Fig. 10 (right) present the metastable density in dependence of the feed gas flow (upper right graph). They were taken in the center of the discharge (2 cm apart from the nozzle in the effluent) while the jet was running surrounded by ambient air. The density increases with the gas flow rate and shows a weak tendency to saturate at higher flows. This strong dependence cannot be explained by a fundamental change in the excitation mechanism, since neither the pressure nor the power coupling to the discharge is changing. To determine the origin of this mechanism, the decay rate of the density in the afterglow was measured. As a result, Fig. 10 right (bottom right graph, orange triangles) indicates that the lifetime of the species increases with the gas flow and, like the density, shows a tendency to saturate for higher flows. Assuming infinite purity of the helium gas and proposing that the metastable lifetime is mainly determined by the three-body-collision process with ground state atoms, the theoretical lifetime of He* ($^3S_1$) in pure helium is calculated to be about 5.7 μs. The loss of metastable atoms by two-body-collisions as well as by diffusion can be omitted since the diffusion loss frequency is 2 orders of magnitude lower than the loss frequency by three-body-collisions. The measured lifetimes in the jet discharge are significantly below this value. The discrepancy with the experimental values can be attributed to the loss of metastable atoms by the Penning ionization process.

Assuming that the dominant impurity contribution is due to residual $N_2$ and $O_2$ molecules, the impurity level can be estimated by a simple model, considering rate coefficient for three-body collisions, quenching and Penning ionizations (from [33, 34]). The result of the model is shown in Fig. 10 b) (bottom right graph, green circles). It agrees very well with the measured values and determines the level of air to be in the order of a few hundred ppm (450 ppm for 1 slm flow, 100 ppm for 5 slm flow). To specify the origin of the impurities, the airtight vessel around the jet was evacuated and backfilled with pure helium. By adding a controlled $N_2$ admixture to the feed gas one could carry out a controlled titration analysis of the impurity level in the jet. These measurements suggest that the intrusion of air can be estimated to be about 75 ppm. Relating this to the results in Fig. 10 (left graph) and considering that the titration analysis was done with an absolute flow of 4 slm, we conclude that roughly 40% of feed gas contamination is caused by impurities in the gas supply system and 60% by the jet being exposed to the ambient air. Operating the jet in ambient air, the decrease of the lifetime can be clearly attributed to back-flow of air through the end of the jet (near the electrodes where the gas velocity profile decays) and to small impurities coming through leakages of the gas feeding system. This is supported by OES and flow simulations (not shown) and can now be quantified by the titration measurements.

**5. O in front of a surface**

Various materials were exposed to the effluent of the jet at different distances and the local O density was measured by TALIF from the exit nozzle of the µ-APPJ up to about 0.1 mm before the targets [35]. The latter distance was limited by spatial access. The schematic arrangement is shown in Fig 11. Atomic oxygen is created in the acitive discharge region of the jet, see Figs. 1 and 4. It is transported in a convection zone close to the target. Note that the carrier gas stream is mostly He. The O concentration decreases along this path by recombination. The flow has a stagnation point at the target surface. It diverts from the longitudinal motion into a flow parallel to the surface in close proximity to the surface. O atoms may reach the target surface by diffusion. This diffusion zone has an estimated width of about 0.1mm.

The flow was modelled with a velocity field calculated following Smith [36]. Measuring the decay profile of O along the x-direction it was concluded, that the O atoms have an effective lifetime of about 42 µs, independent of the gas flow and the starting density at the nozzle, i.e. independent of the RF power coupled to the jet comparable to Jeong [37]. This indicates a constant gas phase chemistry in the free effluent. Measuring the O profiles perpendicular to the jet (at 6 mm from the nozzle) a gaussian shape is found with a half width of 0.6 mm, independent of the gas flow. This indicates that there is no noticeable perpendicular diffusion of O along the path. Of course, the total O intensity increases with flow in agreement with the constant effective lifetime. When polyethylene (PET) and gold are used as target materials at this location a different behaviour of the O concentration in front of the targets is measured (Fig. 12). In the case of (PET), O continues to decay as in the free effluent, whereas an increase is observed in the case of gold.

A simple model involving the surface loss coefficient ß was set up to describe the profiles close to the stagnation points for a fully absorbing wall (particle loss probability ß =1), a fully reflecting surface (ß=0), and an O source at the surface (ß<0). The measured profiles are in accordance with full absorption of O by the PET. O etches PET and the reaction products CO and $CO_2$ are released into the gas phase. Etching of PET was confirmed by measuring the surface topography. These data also show in addition to a localized "deep" etching pattern from O and UV a very broad affected zone which is not associated with the narrow beam particles and radiation. The broad zone may be due to surface attack by $O_3$ which is simultaneously formed in reactions of O with $O_2$ from air.

The experiments with a gold surface indicate, according to the model prediction, O production at the surface (ß<0). This may be explained by a catalytic decomposition of $O_3$ forming O on gold. First ozone measurements using absorption spectroscopy seem to confirm this picture.

## 6. Conclusions

Our analysis of the entire integrated system of a micro plasma jet operated in He/$O_2$ and our attempts to provide quantitative data, have led to significant progress in understanding the behaviour of the complex system from gas introduction to the treated target. We have been able to benchmark model predictions. It became obvious that it is not enough just to quantify the discharge as such (which may be difficult enough) but that it is important to consider carefully all intervening external factors. Nevertheless, although a significant effort has gone into characterizing the system quantitatively, some problems still await resolution.

## Acknowledgement

The work was supported by DFG within the Research Unit FOR 1123.

**Figures:**

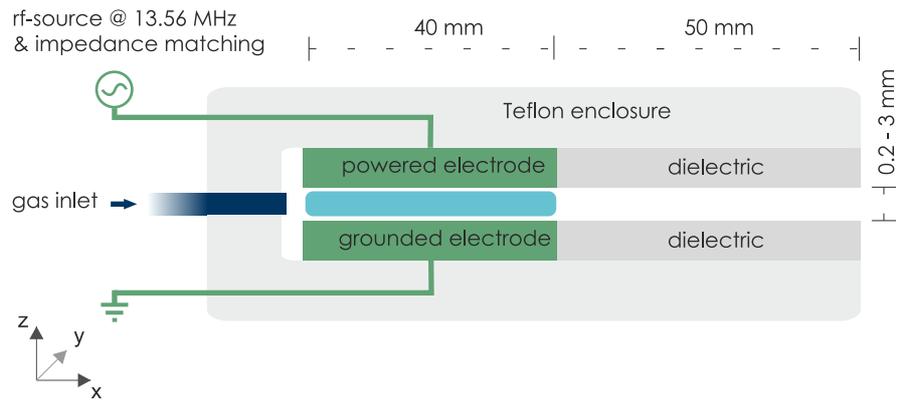

Figure 1: *The µ-APPJ microplasma jet*

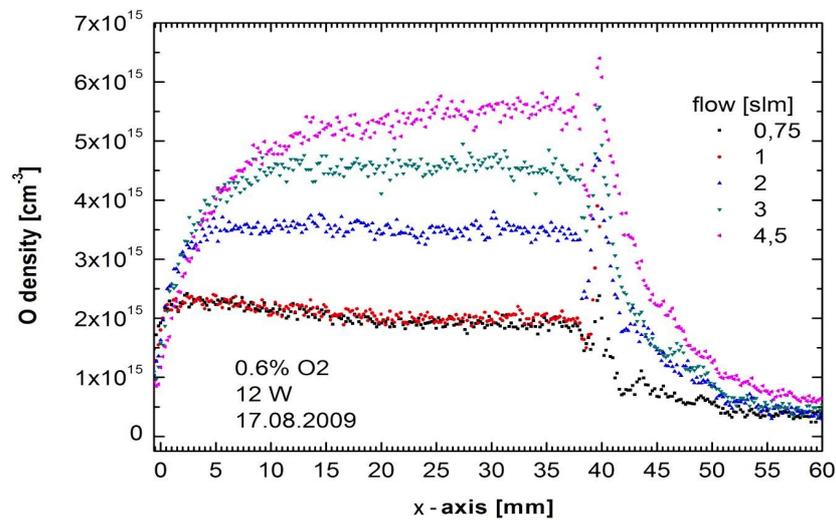

Figure 2: *Absolute O densities in the discharge region and in the effluent of the µ-APPJ*

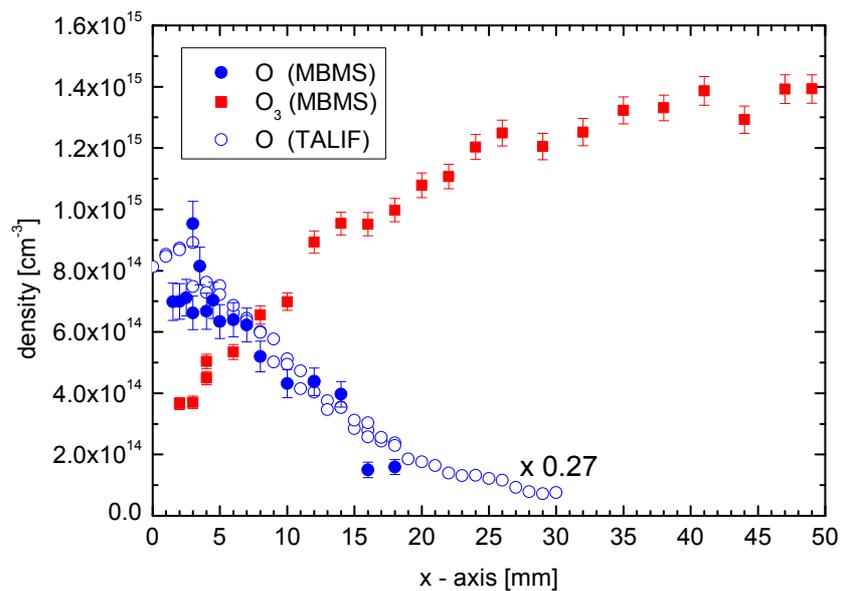

Figure 3: *Comparison of TALIF and MBMS measurements in the effluent of the µ-APPJ*

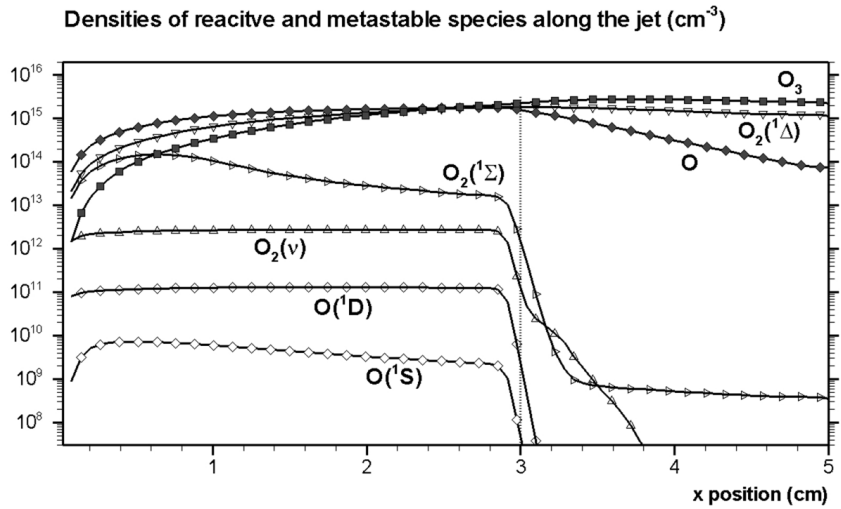

Figure 4: *Model calculations of neutral reactive species*

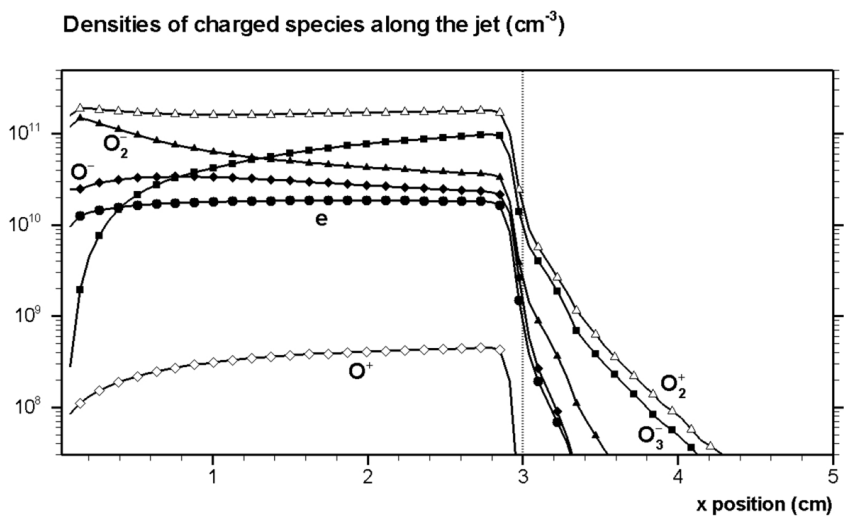

Figure 5: *Model calculations of charged species*

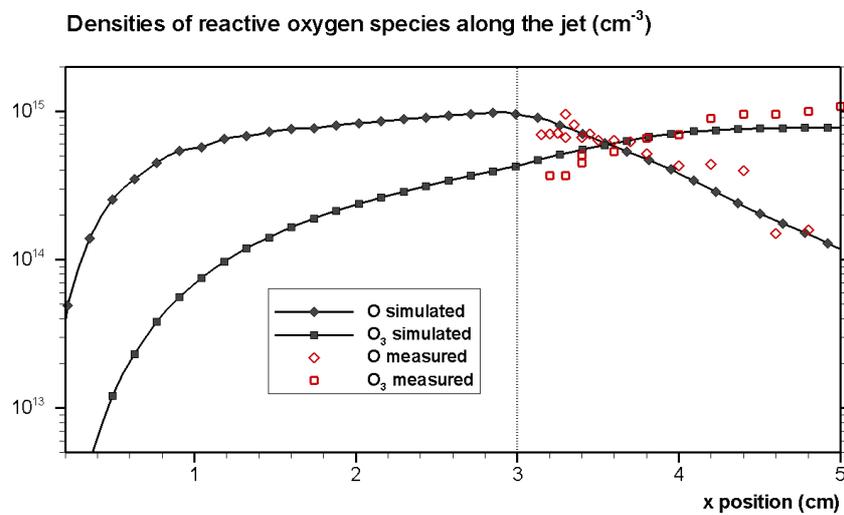

Figure 6: *Comparison of simulation (solid lines) and measurements of O and $O_3$ (symbols) in the µ-APPJ*

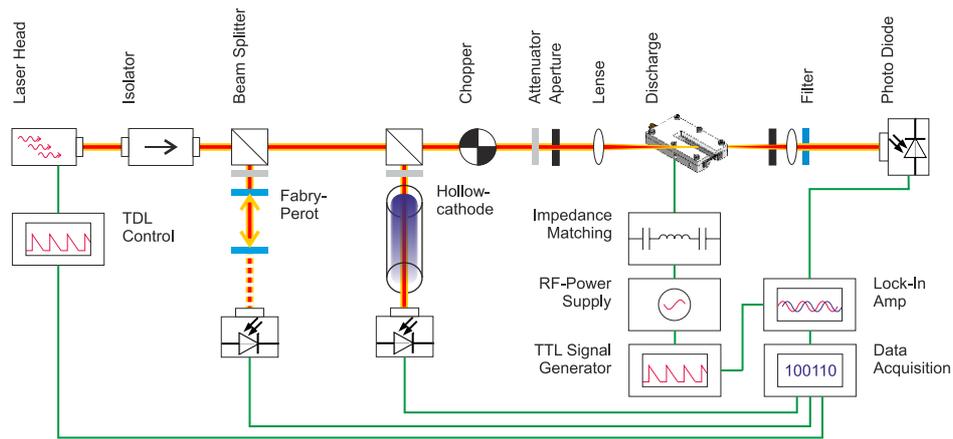

Figure 7: *Experimental setup for TDLAS measurements*

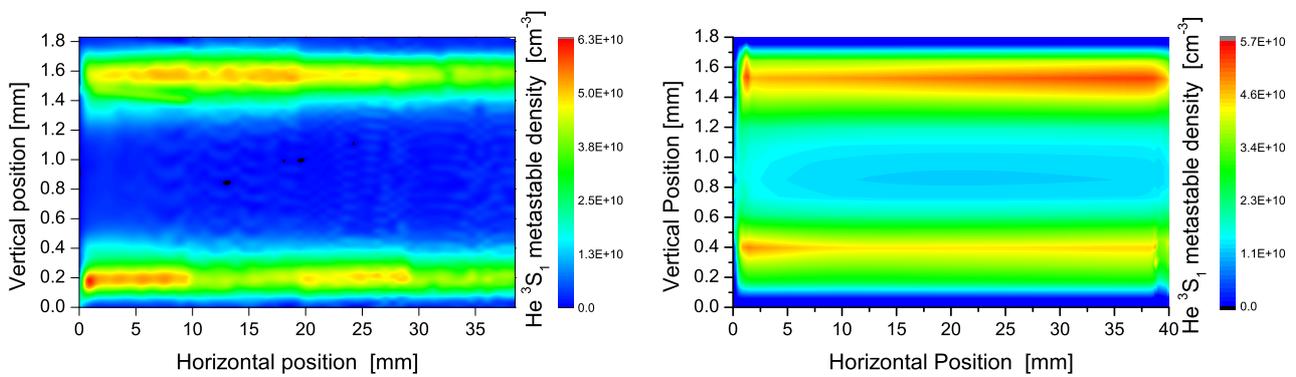

Figure 8: *2d profile of He* metastables in the discharge region of the µ-APPJ measured by TDLAS (left side) and modeling results (right side)*

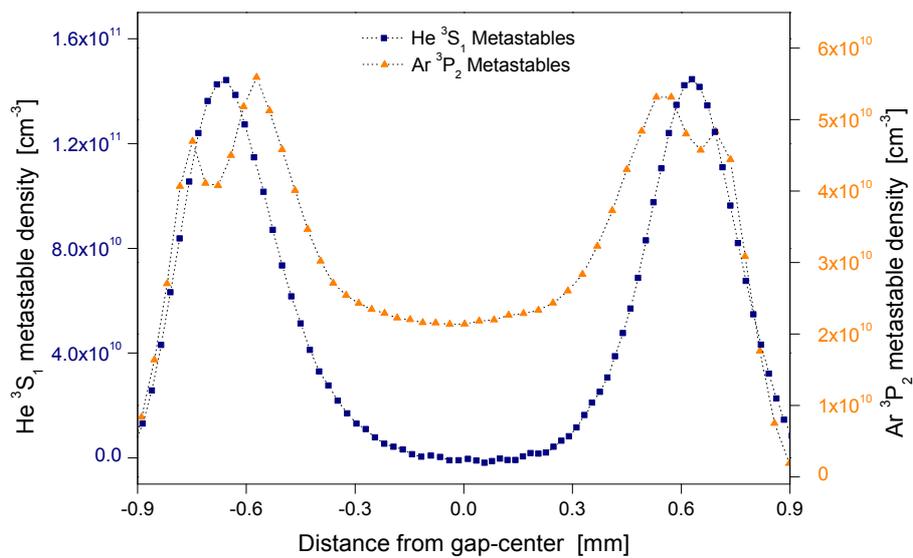

Figure 9: *He* and Ar* metastable profiles across the µ-APPJ*

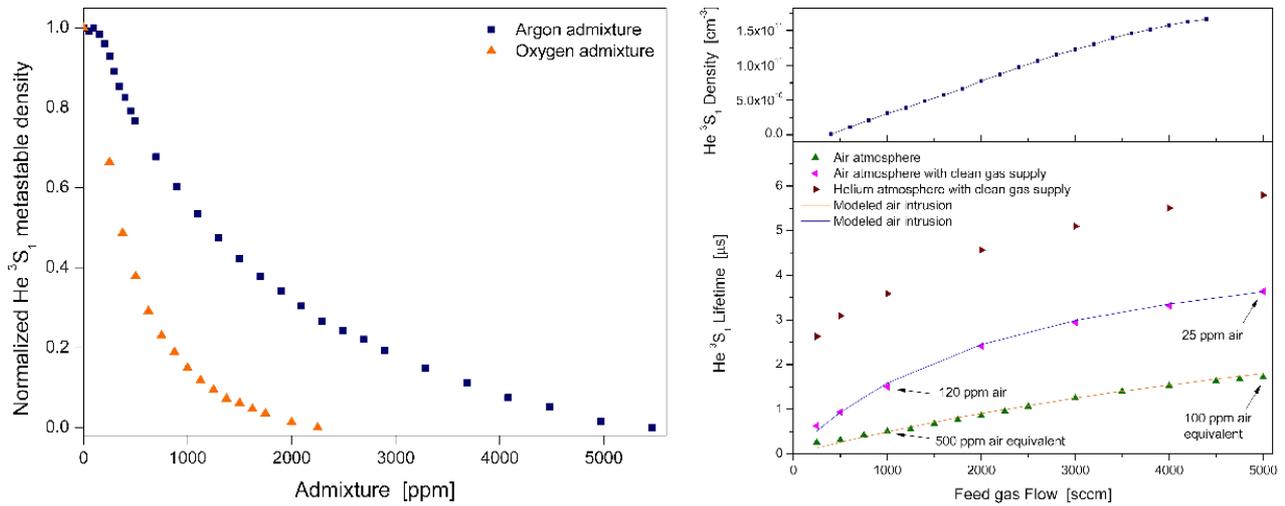

Figure 10: *a Left: Influence of Ar, $N_2$ and $O_2$ admixtures on the metastable density. b Right: He metastable density (top) and lifetime (bottom) in dependence on the feed gas flow. Also shown is the modeled lifetime (orange dots) taking into account significant contribution of nitrogen and oxygen quenching*

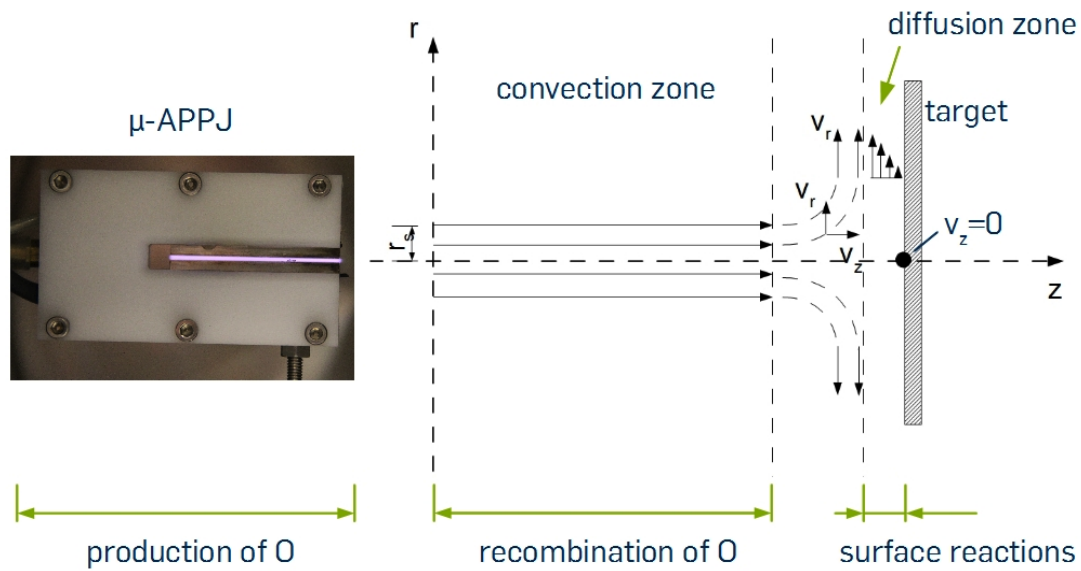

Figure 11: Experimental set*up for measuring the O impinging on a surface*

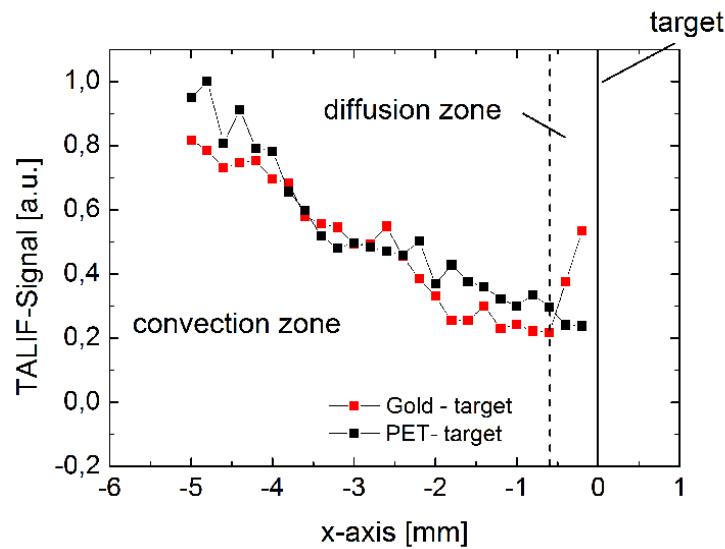

Figure 12: *O profiles in front of targets*